\documentclass[%
 jap,
 amsmath,amssymb,
 reprint,%
]{revtex4-1}

\usepackage{graphicx}
\usepackage{dcolumn}
\usepackage{bm}

\usepackage[utf8]{inputenc}
\usepackage[T1]{fontenc}
\usepackage{mathptmx}
\usepackage{etoolbox}

\makeatletter
\def\@email#1#2{%
 \endgroup
 \patchcmd{\titleblock@produce}
  {\frontmatter@RRAPformat}
  {\frontmatter@RRAPformat{\produce@RRAP{*#1\href{mailto:#2}{#2}}}\frontmatter@RRAPformat}
  {}{}
}%
\makeatother
\begin{document}

\preprint{AIP/123-QED}

\title{Molecular dynamics study of diffusionless phase transformations in HMX: $\beta$-HMX twinning and $\beta$-$\epsilon$ phase transition}
\author{Andrey Pereverzev}%
 \email{pereverzeva@missouri.edu.}
\affiliation{ 
Department of Chemistry and Materials Science $\&$  Engineering Institute, University of Missouri, Columbia, MO 65211, USA
}%

\date{\today}

\begin{abstract}
We use molecular dynamics to study mechanism of deformation twinning of $\beta$-1,3,5,7-tetranitro-1,3,5,7-tetrazocane ($\beta$-HMX) in the $P2_1/n$ space group setting for the twin system specified by $K_1=(101)$, $\eta_1=[10\overline{1}]$, $K_2=(10\overline{1})$, and $\eta_2=[101]$ at $T=1$ K and 300 K. 
Twinning of a single perfect crystal was induced by imposing increasing stress. The following three forms of stress were considered: uniaxial compression along $[001]$, shear stress in $K_1$ plane along $\eta_1$ direction, and shear stress in $K_2$ plane along $\eta_2$ direction. In all cases the crystal transforms to its twin by the same mechanism: as the stress increases, the $a$  and $c$ lattice parameters become, respectively, longer and shorter; soon after the magnitude of $a$ exceeds that of $c$ the system undergoes a quick  phase-transition-like transformation.  This transformation can be approximately separated into two stages: glide of the essentially intact $\{101\}$ crystal planes along $\langle10\overline{1}\rangle$ crystal directions followed by rotations of all HMX molecules accompanied by N-NO$_2$ and CH$_2$ group rearrangements. The overall process corresponds to a military transformation. If uniaxial compression along $[001]$ is applied to a $\beta$-HMX crystal which is already subject to a hydrostatic pressure $\gtrsim 10$ GPa, the transformation described above proceeds through the crystal-plane gliding stage but only minor molecular rearrangements occurs. This results in a high-pressure phase of HMX which belongs to the $P2_1/n$ space group. The coexistence curve for this high-pressure phase and $\beta$-HMX is constructed using the harmonic approximation for the crystal Hamiltonians.

\end{abstract}

\maketitle

\section{Introduction}

The high explosive 1,3,5,7-tetranitro-1,3,5,7-tetrazocane, also known as octahydro-1,3,5,7-tetranitro-1,3,5,7-tetrazocine (HMX), is an important energetic material which is used in a number of high performance military explosive and propellant formulations.\cite{Gibbs} Several polymorphs of HMX are known at standard ambient conditions,\cite{Choi,Kohno,Cady,Cobbledick} among which $\beta$-HMX is the one that is thermodynamically stable. Several other polymorphs were reported, most of them under high pressure.
Yoo and Cynn \cite{Yoo} used Raman spectroscopy and x-ray diffraction to study phase transition of $\beta$-HMX in the quasi-hydrostatic pressure up to 45 GPa. They found evidence of $\beta$-HMX transforming to $\epsilon$-HMX at 12 GPa and $\epsilon$-HMX transforming to $\varphi$-HMX at 27 GPa. Molecular dynamics simulations by Lu et al.\cite{Lu} showed sudden change in lattice parameters of $\beta$-HMX at 27 GPa, which could be related to a phase transition. [17] However, Zhang et al.,\cite{Zhang} using first-principles calculations, did not find evidence of a phase transition at 27 GPa. Korsunskii et al.\cite{Korsunskii} obtained a polymorph belonging to $P2_1/c$ space group with 10 molecules per unit cell; this polymorph is metastable at standard ambient conditions, it quickly transforms to $\beta$ form when ground in a mortar.
Sui et al.\cite{Sui} and Gao et al.\cite{Gao2} used Raman and infrared spectroscopy and x-ray diffraction under hydrostatic and non-hydrostatic pressure. They saw evidence of several phase transitions for pressures between 5 and 40 GPa. Recently, Fan et al.\cite{Fan2} performed simulations of uniaxial compression of $\beta$-HMX using DFT and reported a possible structural transformation at the pressure of 6.82–9.15 GPa for $[100]$ orientation. 

In addition to exhibiting polymorphic transformations, HMX is known to undergo twinning. In particular, $\beta$-HMX can form both growth and deformation twins. Growth twins of $\beta$-HMX with the $(101)$ twinning plane were reported.\cite{Palmer, Heijden, Gallagher1} Several groups  studied plastic deformation of $\beta$-HMX experimenatlly. Palmer and Field investigated compression and indentation of HMX and reported the formation of (101) twins.\cite{Palmer} Cady \cite{Cady2} showed that strained HMX crystals form (101) twins which are reversible upon load removal when created under small strains and irreversible when produced under large strains and at temperatures above 373 K.  Microhardness indentation experiments on $\beta$-HMX conducted in Ref. \cite{Gallagher2} showed evidence of twinning on (101) plane.
 The twin system for deformation twinning of $\beta$-HMX specified by $K_1=(101)$, $\eta_1=[10\overline{1}]$, $K_2=(10\overline{1})$   , and $\eta_2=[101]$ was identified by  Armstrong et al.\cite{Armstrong}
 Pal and  Picu \cite{Pal} used molecular dynamics to study dislocation slip  in $\beta$-HMX for (011) and (101) slip planes and saw evidence of twinning in the case of (101) slip plane. A crystal twinning plasticity model of $\beta$-HMX was developed by Zecevic et al.\cite{Zedevic} Cawkwell et al.\cite{Cawkwell} calculated the energy the (101) twin boundary using DFT approach.
 
The detailed mechanism of the deformation twin formation in $\beta$-HMX is not well understood. Armstrong et al.\cite{Armstrong} proposed a mechanism which involves $\{101\}$ crystal planes translation in $[10\overline{1}]$ direction followed by molecular rotation by 180\textdegree. This mechanism is characterized by high potential energy barrier. In this work we use molecular dynamics to study the twinning mechanism in $\beta$-HMX and show that it involves much more moderate molecular rotation accompanied by some conformational changes.

\section{$\beta$-HMX deformation twinning}
\subsection{Simulation force field}
All molecular dynamics simulations were performed using the LAMMPS package.\cite{Plimpton} We employed the nonreactive, fully flexible molecular force field for HMX proposed by Smith and Bharadwaj \cite{Smith99} and further developed by Bedrov et al.\cite{Bedrov3} and others.\cite{Das,Kroonblawd1} This force field is well-validated and has been used in numerous previous studies of HMX.\cite{Pal,Chitsazi,Perriot2,PereverzevSewell,Bedrov01,Bedrov3,Sewell,Mathew,Bedrov2,Pereverzev2020} We used the version of the force field first described in Ref. \cite{Das}. Sample LAMMPS input decks including all force-field parameters and details of how the forces were evaluated can be found in the supplementary material of Ref. \cite{PereverzevSewell}.

\subsection{Twinning induced by the uniaxial compression along [001]}

To study the mechanism of deformation twining associated with the twin system specified by $K_1=(101)$, $\eta_1=[10\overline{1}]$, $K_2=(10\overline{1})$, $\eta_2=[101]$ we initially performed 
molecular dynamics simulation of uniaxial compression of $\beta$-HMX at 1 K and zero initial pressure. At this temperature the thermal noise is eliminated and the twinning mechanism can be seen clearly.  

The unit cell of $\beta$-HMX in the $P2_1/n$ space group setting is shown in Fig. \ref{Figure2}(a).
The uniaxial compression simulations were performed for a three-dimensionally periodic supercell of $\beta$-HMX comprising $4\times4\times4$ unit cells (3584 atoms): we verified that using a larger supercell ($16\times8\times16$ unit cells) had little effect on the basic twinning transformation steps. 

\begin{figure}
 \includegraphics[width=\columnwidth]{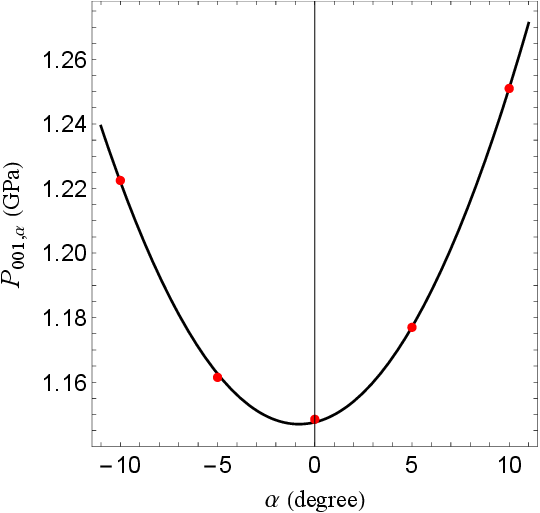}  
 \caption{\label{Figure1} The uniaxial pressure that induces twinning (red circles) as a function of $\alpha$, the angle between the [001] and compression directions in $(010)$ plane. The black curve shows the best fit to a quadratic function. 
}
 \end{figure}

\begin{figure}
 \includegraphics[width=\columnwidth]{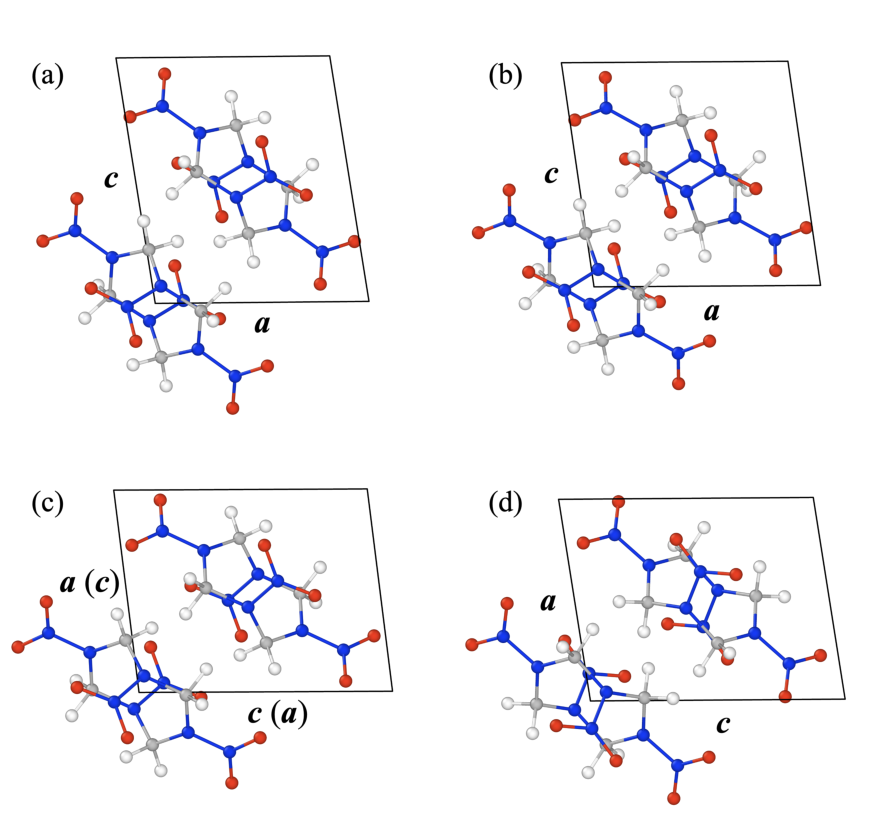}  
 \caption{\label{Figure2} Key steps of $\beta$-HMX deformation twinning due to uniaxial compression along [001] (\textbf{\textit{c}} lattice vector). Projection of the unit cell onto the $(010)$ plane is shown. (a) $\beta$-HMX before compression; (b) compressing the crystals leads to the lattice parameters $a$ and $c$ becoming equal; (c) rapid glide of $\{101\}$ planes leads to switching of \textbf{\textit{a}} and \textbf{\textit{c}} lattice vectors, vectors in parentheses indicate that the current \textbf{\textit{a}} and \textbf{\textit{c}} evolved from the original \textbf{\textit{c}} and \textbf{\textit{a}}, respectively; (d) HMX molecules rotate and their functional groups rearrange to form the twinned $\beta$-HMX unit cell. See text for details. 
}
 \end{figure}

 We  compressed $\beta$-HMX uniaxially along several directions lying in the $(010)$ plane in the vicinity of the [001] direction; this direction was chosen because uniaxial compression along [001] induces shear along both $[10\overline{1}]$ ($\eta_1$) and $[101]$ ($\eta_2$).  The pressure along a given direction was increased from 0 to 2 GPa over the time interval of 1 ns using NPT simulations.
The pressure values for which twinning occurs, $P_{001,\alpha}$, are shown in Fig. \ref{Figure1} as functions of $\alpha$, the angle between the [001] direction  and the compression direction in $(010)$ plane. The minimum of the fitted curve in Fig. \ref{Figure1} lies close to the direction with $\alpha=0$. For this reason, we used uniaxial compression along [001] for further studies of uniaxial compression. 
 
Note that pressure tensor parameters for NPT simulations in LAMMPS are specified with respect to the Cartesian frame and not the crystal frame. As a result,  if the direction of uniaxial compression and a chosen crystal direction are exactly aligned at the start of the simulation they will not, in general, be exactly aligned at a higher pressure because the simulation box deforms under stress. 
  To circumvent this problem we note that LAMMPS always aligns one of the three edges of the simulation box with the \textbf{\textit{x}} Cartesian axis. To ensure that the compression was directed exactly along [001] for any pressure, we aligned $[001]$ direction of the $4\times4\times4$ supercell with the \textbf{\textit{x}} axis and applied compression along \textbf{\textit{x}}. 

\begin{figure}
 \includegraphics[width=\linewidth]{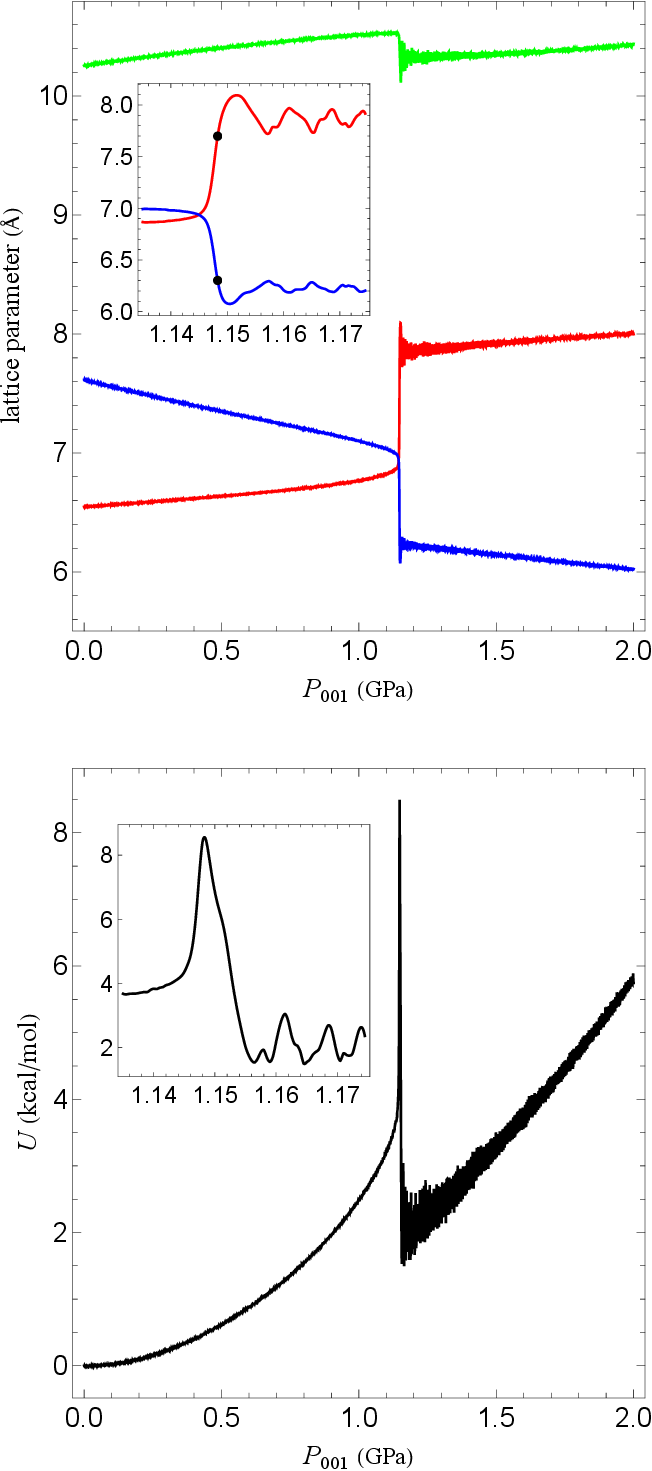}  
 \caption{\label{Figure3} (a) Lattice parameters $a$ (red), $b$ (green), and $c$ (blue)  as functions of $P_{001}$ at $T=1$ K. The inset shows $a$ and $c$ in the vicinity of the glide stage; the two black dots denote $a$ and $c$ values for which the crystal potential energy (see panel (b)) reaches its maximum value.(b) The unit cell potential energy $U$ as a function of $P_{001}$ at $T=1$ K. The inset shows $U$ in the vicinity of the glide stage.     
}
 \end{figure}

The crystal transforms to its twin as the pressure tensor component along [001], $P_{001}$, reaches a critical value.  The transformation is military and follows the mechanism whose main features are shown in Figs. \ref{Figure2} and \ref{Figure3}. 
As the magnitude of the stress increases, the  lattice vectors \textbf{\textit{a}} and \textbf{\textit{c}}  become, respectively, longer and shorter and the system potential energy increases. Fig. \ref{Figure2} (b) shows the compression stage at which the lattice parameters  $a$ and $c$ become equal. Soon after the magnitude of 
\textbf{\textit{a}} exceeds that of \textbf{\textit{c}} (Fig. \ref{Figure3} (a)) the system undergoes a quick, phase-transition-like 
transition to its twin. This quick transition can be approximately separated into two stages: glide of the essentially intact $\{101\}$ crystal planes along 
$\langle10\overline{1}\rangle$ crystal directions (going from Fig. \ref{Figure2} (b) to Fig. \ref{Figure2}(c)) followed by 
rotations of HMX molecules accompanied by N-NO$_2$ and CH$_2$ group rearrangements the most pronounced of which is the equatorial amino nitrogens pyramidal inversion (going from Fig. \ref{Figure2}(c) to Fig. \ref{Figure2}(d)). The separation into the glide and rotation-rearrangement stages is approximate in that the molecular rotation begins before the glide stage is fully complete.
For a given HMX molecule, the molecular rotation axis is formed by intersection of the plane passing through the four carbon atoms and the plane passing through the four nitrogen atoms of the axial N-NO$_2$ groups. The rotation angle when transforming from the structure 
in Fig. \ref{Figure2}(c) to that in Fig. \ref{Figure2}(d) is approximately 24\textdegree.  Overall, the transformation results in the $\beta$-HMX crystal in which, in the Cartesian frame, the original  lattice vectors \textbf{\textit{a}} and \textbf{\textit{c}} are switched 
and the  lattice vector \textbf{\textit{b}} is flipped. The twinned unit cell in Fig.  \ref{Figure2}(d) can be viewed as a reflection of the original unit cell in Fig. \ref{Figure2}(a) in a plane that passes through \textbf{\textit{b}} and bisects $\beta$, the angle formed by \textbf{\textit{a}} and \textbf{\textit{c}}. 
The two unit cells are related through the $P2_1/n$ space group symmetries and, therefore, both represent $\beta$-HMX unit cells. The unit cell potential energy $U$ (Fig. \ref{Figure3}(b)) is an approximately quadratic function of $P_{001}$ for $P_{001} < 1$ GPa;
$U$ reaches its maximum value at the end of the $\{101\}$ planes gliding stage and decreases as the molecules rotate and their 
functional groups rearrange. The value of $P_{001}= 1.1483\pm0.0001$ GPa which induces the twinning was taken to be the value at which 
the derivatives of lattice parameters $a$ and $c$ with respect to $P_{001}$ have the highest magnitude. A video of the twinning transformation is provided in the supplementary material.  

Applying uniaxial compression along [001] to the same $4\times4\times4$ supercell of $\beta$-HMX at $T$= 300 K and zero initial pressure induces deformation twinning that follows the same basic steps as at $T$=1 K. Apart from the presence of thermal noise, the main difference from the 1 K case is that the value of $P_{001}$ at which the twinning occurs is lower and given by $0.792\pm0.004$ GPa.
\subsection{Twinning induced by shear} \label{shear}

The twin system studied here is specified by the conjugate twinning planes $K_1=(101)$, $K_2=(10\overline{1})$ and conjugate twinning directions $\eta_1=[10\overline{1}]$ and $\eta_2=[101]$. This implies that the deformation twinning can also be induced by 
applying shear stress either in $K_1$ plane along $\eta_1$ direction or in $K_2$ plane along $\eta_2$ directions. We studied both cases at $T$=1 K and 300 K. Evaluation of the (101) twin boundary energy is given in the Appendix.

As in the case of the uniaxial compression discussed in the previous subsection, we needed to choose the proper orientation of the simulation box to ensure that the shear stress is always  aligned  with the corresponding twinning plane and twinning direction as the box deforms under stress. 
This could not  be done exactly with the supercell that we used for the uniaxial compression. 
To overcome this, we used a non-conventional 4-molecule unit cell of $\beta$-HMX whose lattice 
vectors ${\textbf{\textit{a}}}'$, ${\textbf{\textit{b}}}'$, and ${\textbf{\textit{c}}}'$  are expressed through the original \textbf{\textit{a}}, \textbf{\textit{b}}, and \textbf{\textit{c}} as ${\textbf{\textit{a}}}'={\textbf{\textit{a}}}-\textbf{\textit{c}}$, ${\textbf{\textit{b}}}'= \textbf{\textit{b}}$, and ${\textbf{\textit{c}}}'={\textbf{\textit{a}}}+\textbf{\textit{c}}$, 
thus the faces of this new unit cell are parallel  to $(101)$, $(010)$, and $(10\overline{1})$ planes in the original $P2_1/n$ space group setting. 

\begin{figure}
 \includegraphics[width=\columnwidth]{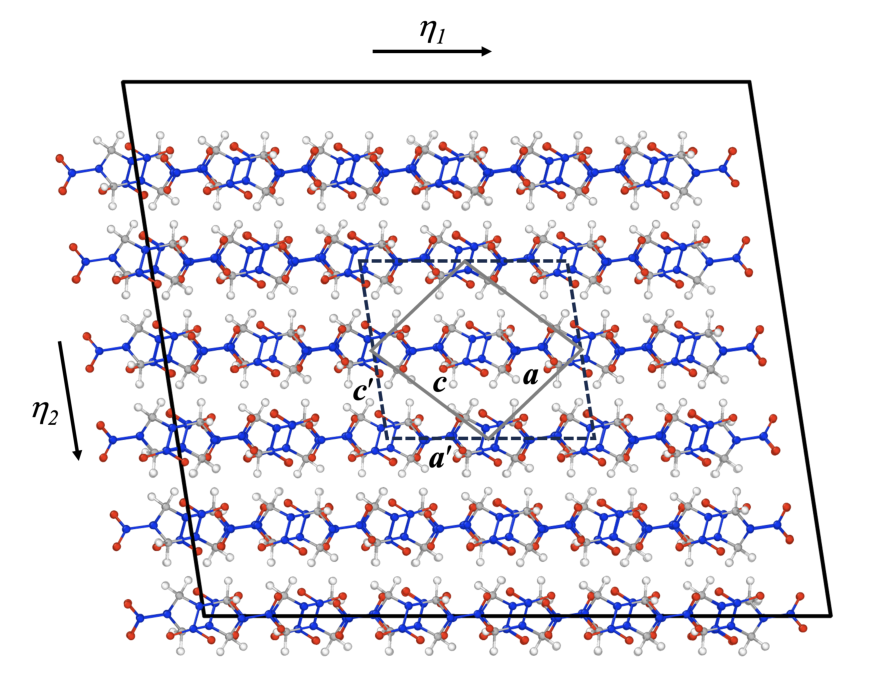}  
 \caption{\label{Figure4} The supercell of $\beta$-HMX used to simulate twinning induced by shear, projected onto the (010) plane. The shear stress is applied either in (101) ($K_1$) along $[10\overline{1}]$ ($\eta_1$) or in $(10\overline{1})$ ($K_2$) along [101] ($\eta_2$). Projections of the conventional $P2_1/n$ unit cell (solid gray) and the non-conventional 4-molecule unit cell (dashes) are also shown. 
}
 \end{figure}

The supercell comprising $3\times4\times3$ of these unit cells (4032 atoms), which was used to study twinning under shear, is shown in Fig. \ref{Figure4}. 
The twinning was induced by increasing the magnitude of the corresponding shear stress component from 0 to 0.8 GPa over the time interval of 1 ns using NPT simulaions. The twinning mechanism both for shear applied in $K_1$ plane and $\eta_1$ direction and in $K_2$ plane and $\eta_2$ direction follows the same basic steps as in the case of the uniaxial compression. The magnitude of the shear stress that induces twinning for the case of $K_1$ and $\eta_1$ is $0.4511\pm0.0001$ GPa at 1 K and $0.320\pm0.002$ GPa at 300 K. For $K_2$ and $\eta_2$ the values are $0.4463\pm0.0001$ GPa at 1 K and $0.318\pm0.002$ GPa at 300 K. Note that for a given temperature, the magnitudes of twinning-inducing stress for the two twinning directions are almost the same. This can be expected: as in the case of the uniaxial compression, twinning occurs in the vicinity of the cell geometry for which lattice parameters $a$ and $c$ are equal or, equivalently, when the simulation cell shown in Fig. \ref{Figure4} becomes orthogonal; if an orthogonal cell is obtained by applying stress either in $K_1$ along $\eta_1$ or in $K_2$ along $\eta_2$, then the magnitudes of these two stresses have to be identical.

\section{$\beta$-$\epsilon$ phase transition} \label{sec3}
\begin{figure}
 \includegraphics[width=\columnwidth]{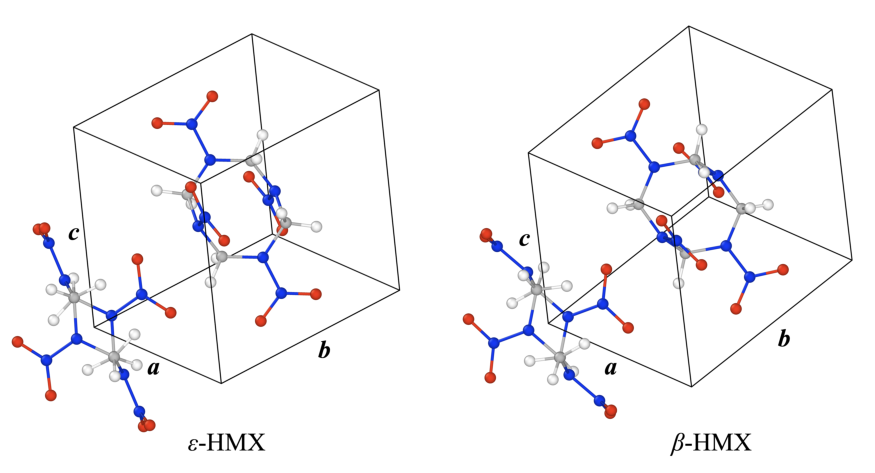}  
 \caption{\label{Figure5} Units cells of $\epsilon$-HMX (left) and $\beta$-HMX (right). Note different orientation and geometry of the HMX molecules.
}
 \end{figure}

 The 
 twinning mechanism discussed in the previous section involves a glide of crystal planes followed by rotation and rearrangement of HMX molecules. This two stage process hints at the possibility of a transformation in which, under proper conditions, the molecular rotation does not occur.
Indeed, if the uniaxial compression along $[001]$  is applied to a $\beta$-HMX crystal which is already subject to a hydrostatic pressure $\gtrsim 10$ GPa, 
the transformation described above proceeds through the crystal-plane gliding stage (Fig. \ref{Figure2}(c)) but only minor molecular rearrangements occur. This results in a high-pressure phase of HMX belonging to the $P2_1/n$ space group that remains stable or metastable for hydrostatic pressure $\gtrsim 10$ GPa (Fig. \ref{Figure7}). For notational  convenience we refer to this phase as $\epsilon$-HMX. Since the transformation from $\beta$- to $\epsilon$-HMX is military, the unit cell of $\epsilon$-HMX consists of two HMX molecules as shown in Fig. \ref{Figure5}. 
The $\beta$ and $\epsilon$ polymorphs differ by molecular orientation, molecular geometry, and the lattice parameter values. The lattice parameters of $\epsilon$-HMX at $T=300$ K and $P=12$ GPa are given by $a=5.9529$ \AA , $b=9.0800$ \AA, $c=7.5720$ \AA, 
$\beta=97.32${\textdegree}  with the unit cell volume $V_{\text {uc}}=405.947$ \AA$^3$. The lattice parameters of $\beta$-HMX under the same conditions are $a=6.2303$ \AA, $b=9.3231$ \AA, $c=7.0413$ \AA, $\beta=97.12${\textdegree} with $V_{\text {uc}}=405.849$ \AA$^3$. Molecules of HMX in the $\epsilon$ polymorph are in the chair conformation but have a somewhat different orientation in the unit cell compared to the $\beta$ form. Also note that the equatorial amino nitrogen atoms exhibit pyramidal inversion with respect to the same atoms in the $\beta$ polymorph. Data files for the unit cells shown in Fig. \ref{Figure5} and a video of the $\beta-\epsilon$ transformation under uniaxial compression are provided in the supplementary materials.

The phase diagram for $\beta$- and $\epsilon$-HMX in $P$-$T$ space was obtained using the condition $\Delta G = 0$,
where $\Delta G$ is the difference of the Gibbs free energies for $\beta$ and $\epsilon$ phases. $\Delta G$ was calculated from
\begin{equation}
\Delta G=\Delta U+P\Delta V-T\Delta S, \label{Eq1}
\end{equation}
where $\Delta U$, $\Delta V$, and $\Delta S$ are, respectively, the differences of internal energies, volumes, and entropies for the two phases. 
The first two terms of the right-hand side of Eq. (\ref{Eq1}) were obtained as functions of pressure and temperature from MD simulations: for a selected fixed temperature NPT simulations were run with increasing hydrostatic pressure for $\beta$- 
and $\epsilon$-HMX. Temperature was sampled every 100 K in the range $0 \leq T \leq 500$ K, for each temperature value the pressure range $12 \leq P \leq 70$ GPa was considered.

$T\Delta S$ term in Eq. (\ref{Eq1}) was calculated using the harmonic approximation for crystal Hamiltonians. For a  three-dimensional periodic harmonic crystal, the classical canonical partition function can be calculated exactly. This partition function can be used to calculate entropy.\cite{Huang}  It can be shown that 
\begin{equation}
T\Delta S=kT\sum_{i=1}^{3N-3} \ln \frac{\omega_i^{\epsilon}}{\omega_i^{\beta}}, \label{Eq2}
\end{equation}
where $\omega_i^{\epsilon}$ and $\omega_i^{\beta}$ are the normal mode frequencies for $\epsilon$- and $\beta$-HMX, respectively, and $N$ is the number of atoms in the system. The 
frequencies $\omega_i^{\epsilon}$ and $\omega_i^{\beta}$ as functions of temperature and pressure were calculated using the normal mode analysis. For given pressure and temperature the crystal potential energy was minimized keeping simulation box parameters fixed, second derivatives of potential energy with respect to atomic displacements were calculated and the corresponding dynamical matrix was diagonalized to obtain normal mode frequencies; the detailed description of this procedure can be found in Ref. \cite{Pereverzev2011}. Similar calculations of pressure-dependent normal mode frequencies were done for 
molecular crystals of PETN and RDX in Refs. \cite{Pereverzev2011b,Pereverzev2013}.  

\begin{figure}
 \includegraphics[width=\columnwidth]{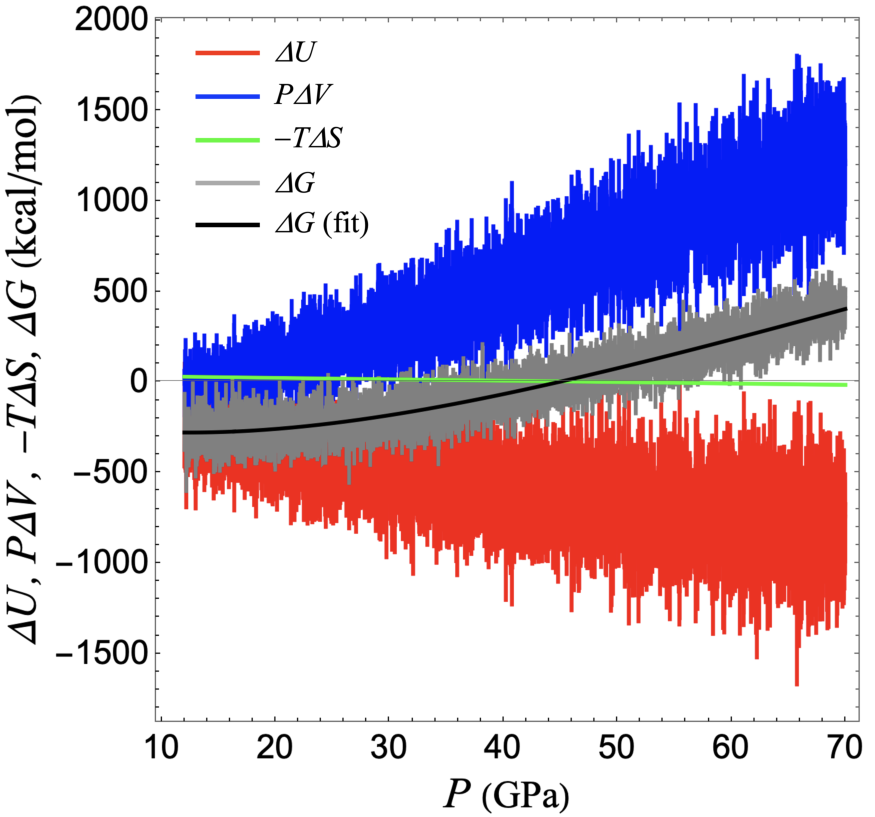}  
 \caption{\label{Figure6} $\Delta U$ (red), $P\Delta V$ (blue), $-T\Delta S$ (green), and $\Delta G$ (gray) as functions of pressure at 300 K. The black curve shows the best fit of a quartic polynomial to the $\Delta G$ data. 
}
 \end{figure}

\begin{figure}
 \includegraphics[width=\columnwidth]{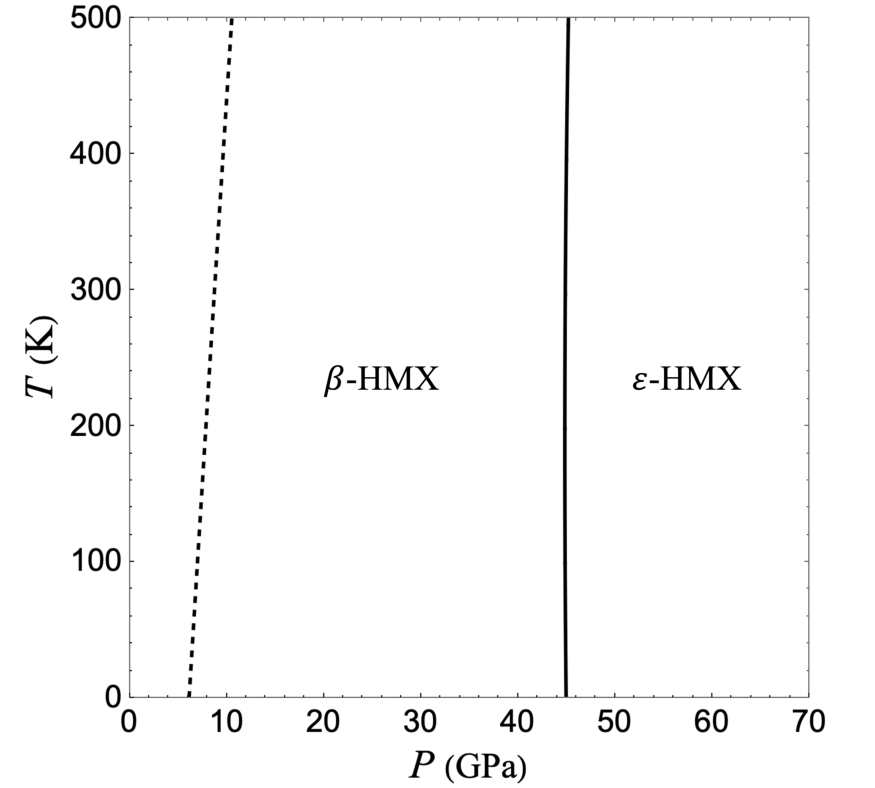}  
 \caption{\label{Figure7} Phase diagram of $\epsilon$- and $\beta$-HMX coexistence region. The phase boundary is shown in solid black. $\epsilon$-HMX is metastable in the region between the dashed curve and the phase boundary.
}
 \end{figure}

As an example, Fig. \ref{Figure6} shows $\Delta G$ and the constitutive $\Delta U$, $P\Delta V$, and $-T\Delta S$ as functions of pressure at $T=300$ K. Note that the contribution of $T\Delta S$ term to $\Delta G$ is small in comparison to $\Delta U$ and $P\Delta V$ terms. A fourth-degree polynomial was fitted to the raw data for $\Delta G(P)$. This polynomial was then used to 
find the pressure at which $\Delta G(P)=0$ for a given temperature. 
 The resulting phase diagram is shown in Fig. \ref{Figure7}. $\epsilon$-HMX is thermodynamically stable above approximately 45 GPa; the phase boundary shows very weak dependence on temperature over the considered temperature range. $\epsilon$-HMX remains metastable well below 45 GPa. The dashed curve in Fig. \ref{Figure7} corresponds to the pressure and temperature at which $\epsilon$-HMX transforms to $\beta$-HMX. In particular, $\epsilon$-HMX transforms to $\beta$ polymorph at approximately 6.1 GPa at $T=1$ K, 8.7 GPa at $T=300$ K, and 10.5 GPa at $T=500$ K; the transformation is military.  A video of $\epsilon$-$\beta$ transformation is provided in the supplementary materials.
 Note that the phase diagram in Fig. \ref{Figure7} does not include $\alpha$, $\delta$, or other possible high-pressure polymorphs of HMX.
 \section{Discussion}
 Armstrong et al.\cite{Armstrong} studied deformation twinning of $\beta$-HMX  for the twin system specified by $K_1=(101)$, $\eta_1=[10\overline{1}]$, $K_2=(10\overline{1})$   , and $\eta_2=[101]$.  
 They proposed a twinning mechanism which involves $\{101\}$ crystal planes translation in $[10\overline{1}]$ direction followed by molecular rotation by 180\textdegree. This mechanism is characterized by a high potential energy barrier. Our molecular dynamics results show that 
 $\beta$-HMX deformation twinning does involve $\{101\}$ crystal planes translation and molecular rotation; however, the molecular rotation that leads to a twin structure is much more moderate compared to the one proposed by  Armstrong et al. At the same time, the mechanism that we observe in this study involves a change of molecular conformation: each HMX molecule transforms to its mirror image. The twinning mechanism for crystals with a basis which is qualitatively similar to the one observed in this work was suggested by Bilby and Crocker.\cite{Bilby}

Twinning of $\beta$-HMX studied here may have implications for simulation of shock propagation in  $\beta$-HMX. Our results suggest that shocking $\beta$-HMX along [001] or its vicinity is likely to induce twinning by the same mechanism. A critical aspect of simulating such shocks is to use boundary conditions which would allow  the system to expand in the directions normal to the shock direction.

 Our observation of a high-pressure phase of HMX belonging to $P2_1/n$ space group ($\epsilon$-HMX) finds some support in the experimental results of 
 Sui et al.\cite{Sui} and Gao et al.\cite{Gao2} who reported formation of several phases of HMX under high pressure. Gao et al. identified two of these phases as belonging to  space group $P2_1/c$ (or, equivalently, $P2_1/n$) with two HMX molecules per unit cell. They reported fractional coordinates and lattice parameters for these phases. Unfortunately, we were unable to reproduce meaningful unit cell structures from the fractional coordinates data of Ref. \cite{Gao2} and, therefore, could not directly compare the structure of $\epsilon$-HMX studied in this work to the two phases discussed by Gao et al.

\section{Conclusions}

We used molecular dynamics simulations to clarify the mechanism of twinning in $\beta$-HMX for the twin system specified by $K_1=(101)$, $\eta_1=[10\overline{1}]$, $K_2=(10\overline{1})$, and $\eta_2=[101]$. Our results apply to a single perfect crystal in which twinning proceeds militarily. Twinning in crystals with defects is likely to follow  more complicated paths \cite{Pal} which require additional studies. Molecular dynamics studies of other possible twinning planes and directions in $\beta$-HMX such as the ones discussed in Ref. \cite{Gallagher2} can be a subject of future research.
It is also of interest to compare the molecular dynamics results for a single perfect crystal reported here to predictions of the twinning plasticity model developed by Zecevic et al.\cite{Zedevic}

Our simulations show that the formation of the high pressure phase of $\epsilon$-HMX can be viewed as the twinning transformation that did not proceed to completion. In this sense, the new polymorph appears to be physically plausible. However, additional studies of $\epsilon$-HMX such as high-level quantum calculation would be useful to ensure that it does represent a physical high-pressure phase of HMX. 
 \section{Supplementary Material}
 Supplementary material includes video files of $\beta$-HMX twinning, $\beta$-$\epsilon$ transformation, and $\epsilon$-$\beta$ transformation. LAMMPS data files for $\epsilon$- and $\beta$-HMX unit cells are also provided.

\begin{acknowledgments}
This research was funded by Air Force Office of Scientific Research, Grant No. FA9550-19-1-0318 and by an AFOSR DURIP equipment award, Grant No. FA9550-20-1-0205. The author is grateful to Tommy Sewell for valuable comments.
\end{acknowledgments}

\section*{Data Availability Statement}

The data that support the findings of this study are available from the corresponding author upon reasonable request.
\appendix*
 \section{Energy of the (101) twin boundary}
\begin{figure}
 \includegraphics[width=\columnwidth]{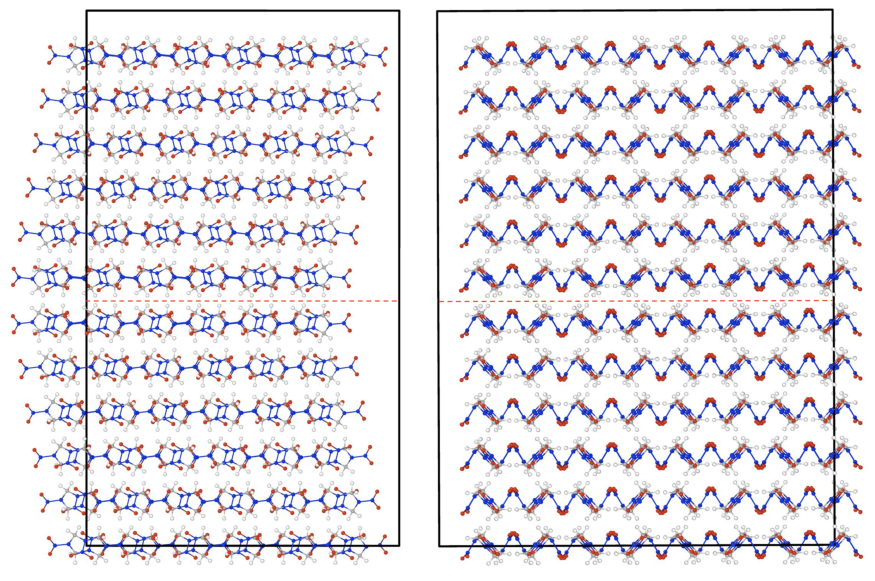}  
 \caption{\label{Figure8} The supercell of $\beta$-HMX, composed of two supercells shown in Fig. \ref{Figure4} and used to estimate (101) twin boundary energy, viewed along
 $[010]$  (left) and $[10\overline{1}]$  (right).
 The twin boundary is shown with red dashes.}
 \end{figure}

It is of interest to estimate the value of the (101) twin boundary energy and compare it to the DFT-calculated results 
obtained by Cawkwell et al.\cite{Cawkwell}. The supercell with (101)  twin boundary  was created using two supercells used in the shear simulations in subsection \ref{shear} and shown in Fig. (\ref{Figure4}). One of these supecells was rotated by 180{\textdegree} about the normal to (101) plane and superposed on the non-rotated supercell to form a twin boundary as shown in Fig.\ref{Figure8}.
Note that, at the twin boundary of the composite supercell, the crests and troughs of the zigzag molecular patterns as seen on the right panel of Fig. \ref{Figure8} are matching for the top and bottom constitutive supercells; this is of relevance to the discussion below. 

The  supercell with  the twin boundary was equilibrated at $T=1$ K and zero pressure using NPT simulations. The average potential energy $E$ of the supercell was calculated from the  500 ps production NPT trajectory.
An identical calculation was performed for a perfect crystal of $\beta$-HMX constructed from $3\times 4\times 6$ non-conventional 4-molecule unit cells discussed in subsecton \ref{shear} to obtain the average potential energy $E_0$ for a crystal without the twin boundary. 
Following Ref. \cite{Cawkwell} the twin boundary energy $\gamma$ was calculated using 
\begin{equation}
    \gamma=\frac{E-E_0}{2 A},
\end{equation}
where $A$ is the area of the twin boundary and the factor of two accounts for the presence of two twin boundaries in a periodic cell.

The resulting energy of the (101) twin boundary is 28.2 mJ/m\textsuperscript{2}. Note that our results apply to a system at $T=1$ K and zero pressure. However, at this temperature, we can expect them to be close to the values obtained by energy minimization, the approach used by Cawkwell et al.\cite{Cawkwell} The value obtained by Cawkwell et al. is 163 mJ/m\textsuperscript{2}, which is approximately six times higher than our result. It seems unlikely that such a big difference in $\gamma$ values is due to  the different computational approaches used. It is  possible that the (101) twin boundary studied by Cawkwell et al. is microscopically different from the one considered here. Cawkwell et al. did not  
provide detailed description or images of the twin boundary that they studied. However, based on their description of the supercell preparation procedure, it appears   that the top half of their supercell is the mirror image of the bottom half. If this is indeed the case then the twin boundary of Cawkwell et al. is different from ours and can be obtained from the one shown in Fig. \ref{Figure8} by 
displacing the top half of the supercell half a unit cell along [010]. This will result in a supercell in which the crests of the zigzag  patterns  for the bottom half of the supercell (as seen on the right panel of Fig. \ref{Figure8}) are facing the troughs (with the same functional groups) of the top half. Such twin boundary has higher energy than the one studied by us and cannot be simulated using the approach used here: the supercell with such twin boundary quickly transforms into the one in Fig. \ref{Figure8} in equilibrium simulations. 


%

\end{document}